\documentclass[]{bmcart}
\usepackage{amsthm,amsmath}
\usepackage{bm}
\usepackage[utf8]{inputenc} 
\usepackage{graphicx}
\graphicspath{{./Figs/}}
\usepackage{cancel}
\usepackage{booktabs}

\usepackage{upgreek}
\usepackage{graphicx}
\usepackage{subfigure}
\usepackage{color}
\usepackage{color}
\usepackage[normalem]{ulem} 
\newcommand{\Replace}[2]{\bgroup\noindent\textcolor{red}{\xout{#1} #2}\egroup\ignorespacesafterend}
\newcommand{\Delete} [1]{\bgroup\noindent\textcolor{red}{\xout{#1}}\egroup\ignorespacesafterend}
\newcommand{\Insert} [1]{\bgroup\noindent\textcolor{}{#1}\egroup\ignorespacesafterend}
\newcommand{\Comment}[1]{\definecolor{Mygray}{gray}{0.50}\bgroup\color{Mygray}\noindent#1\egroup\ignorespacesafterend}
\newcommand \Michael [1]{\bgroup\noindent[\textcolor{blue}{\textbf{Michael}: #1}]\egroup\ignorespacesafterend}
\newcommand \Stefan  [1]{\bgroup\noindent[\textcolor{blue}{\textbf{Stefan}: #1}]\egroup\ignorespacesafterend}

\DeclareMathAlphabet{\Ibb}{U}{msb}{m}{n}

 \newcommand{\dx}{\text{\rm d}x}
 \newcommand{\dy}{\text{\rm d}y}

\newcommand{\taue}{\tau_{\rm ext}}

\newcommand{\erf}{\textrm{erf}}

\newcommand \MZ [1] {\bgroup\noindent[\textcolor{blue}{\textbf{MZ}: #1}]\egroup\ignorespacesafterend}

\begin{document}

\begin{frontmatter}

\begin{fmbox}
\dochead{Research}

\title{Pinning of dislocations in disordered alloys: Effects of dislocation orientation}

\author[
   addressref={aff2},
   corref={aff2},                       
   email={michael.Zaiser@fau.de}
]{\inits{MZ}\fnm{Michael} \snm{Zaiser}}
\author[
addressref={aff3},                 
]{\inits{RW}\fnm{Ronghai} \snm{Wu}}

\address[id=aff2]{
\orgname{Department of Materials Simulation, WW8-Materials Simulation, Friedrich-Alexander Universität Erlangen-Nürnberg}, 
  \street{Dr.-Mack-str. 77},                     %
  \postcode{90762}                                
  \city{Fürth},                              
  \cny{Germany}                                    
}
\address[id=aff3]{
	\orgname{School of Mechanics, Civil Engineering and Architecture, Northwestern Polytechnical University}
	\postcode{710129}                                
	\city{Xian},                              
	\cny{P.R. China}                                    
}

\begin{artnotes}

\end{artnotes}

\end{fmbox}


\begin{abstractbox}

\begin{abstract} 
The current interest in compositionally complex alloys including so called high entropy alloys has caused renewed interest in the general problem of solute hardening. It has been suggested that this problem can be addressed by treating the alloy as an effective medium containing a random distribution of dilatation and compression centers representing the volumetric misfit of atoms of different species. The mean square stresses arising from such a random distribution can be calculated analytically, their spatial correlations are strongly anisotropic and exhibit long-range tails with third-order power law decay  \cite{geslin2021microelasticityI,geslin2021microelasticityII}. Here we discuss  implications of the anisotropic and long-range nature of the correlation functions for the pinning of dislocations of arbitrary orientation. While edge dislocations are found to follow the standard pinning paradigm, for dislocations of near screw orientation we demonstrate the co-existence of two types of pinning energy minima. 
\end{abstract}


\end{abstractbox}

\end{frontmatter}
´´
\section[Introduction]{Introduction}

The theory of dislocations interacting with atomic-scale obstacles, traditionally formulated in the context of solution hardening, has seen a renaissance in recent years which has been driven by the general interest in compositinally complex alloy systems including so-called high-entropy alloys. In such alloys, multiple atomic species are present in comparable concentrations and entropic effects may stabilize homogeneous phases at elevated temperatures, whereas kinetic effects (slow diffusion due to multiple barriers and traps) may stabilize those phases against unmixing at reduced temperatures. 

From a theoretical viewpoint, statistical theories of dislocation pinning by atomic-scale obstacles have, starting from the seminal work of Labusch \cite{Labusch1970_PSS,Labusch1972_AM}, attracted the interest of statistical physicists, and concepts developed for the pinning of elastic manifolds by random fields (e.g. \cite{Chauve2000_PRB}) and their depinning by external forces were applied to the athermal motion of dislocations (e.g. \cite{Zapperi2001_MSEA,Bako2008_PRB}) and to dislocation motion at finite temperatures \cite{Ioffe1987_JPC, Zaiser2002_PM}. In recent years, these concepts have been extended and adopted to compositionally complex and high entropy alloys by a number of authors (e.g.\cite{Toda2015_AM,Wu2016_AM,Varvenne2016_AM,Varvenne2017_AM,larosa2019solid}). In particular, the group of W. Curtin has demonstrated that pinning of dislocations in compositionally complex alloys and the associated flow stress increase can to a large extent be explained in terms of the significant local stress fluctuations introduced by the superposition of misfit strains associated with atomic species of significantly different atomic radius. Recently,  \cite{geslin2021microelasticityI,geslin2021microelasticityII} evaluated the magnitude and spatial correlations of such random stress/strain fields. Here we use their result to study the effects of dislocation orientation on dislocation pinning by volumetric misfit fluctuations.

\section{Scaling theory of elastic lines in static random fields}
\label{sec:2} 

We envisage the dislocation as an elastic line of line tension ${\cal T}= \beta_0\mu b^2$ where $\mu$ is the shear modulus, $b$ the length of the Burgers vector, and the numerical parameter $\beta_0$ which is of the order of 1 may depend logarithmically on geometrical parameters characterizing the line shape. The dislocation is assumed macroscopically straight while local fluctuations of the line shape are described by a function $y(x)$ where the $x$ axis is oriented along the average line direction and the $y$ axis in perpendicular direction within the glide plane. Note that we assume $y(x)$ to be single valued, which excludes the presence of overhangs. 

Let atomic disorder create a spatially fluctuating but temporally fixed resolved shear stress field $\tau(x,y)$ that acts on the dislocation, which is in addition subject to a spatially constant resolved shear stress $\taue$. The evolution of the dislocation line shape is then given by the quenched Edwards-Wilkinson equation 
\begin{equation}
	B \frac{d y}{d t} = {\cal T}\frac{\partial^2 y}{\partial x^2} + b [\tau(x,y) + \taue]
	\label{eq:qedwilk}
\end{equation}
The random shear stress $\tau(x,y)$ has the correlation function
\begin{equation}
	\langle \tau(x,y)\tau(x',y') \rangle = \langle \tau^2 \rangle  \Phi\left(\frac{|x-x'|}{a},\frac{|y-y'|}{a}\right).
\end{equation}
For a random alloy, the magnitude $\langle \tau^2 \rangle$ can be related to the average of the squared atomic misfit (see below) \cite{geslin2021microelasticityI}. Analytical expressions that allow to compute the correlation function $\Phi$ have been given in Ref.  \cite{geslin2021microelasticityII} where the length scale parameter $a$ was, by comparison with MD results, determined to be of the order of 1\AA. 
For later use we introduce the notations $\Phi_{\perp}(y) := \Phi(x=0,y)$ for the correlation function in the line-perpendicular and $\Phi_{\parallel}(x):= \Phi(x,y=0)$ for the correlation function in the line-parallel direction.

We now evaluate the work per unit length done by the fluctuating stress as an infinitesimal dislocation segment at $(x,0)$ displaces from $y=0$ to $y=w$. This is given by $b \int_0^w \tau(x,y) \dy$. We express the average value of this integral in terms of the conditional average of $\tau(x,y)$ given the stress at $(x,0)$, i,e, $\langle\tau(x,y)\rangle_{\tau(x,0)} = \tau(x,0)\Phi_{\perp}(y/a)$: $\langle W_{\tau}(x,w) \rangle_{\tau(x,0)} = b \tau(x,0) \int_0^w \Phi_{\perp}(y/a) \dy$. 

Next we consider a segment of finite length $L$ which is displaced by a distance $w$ in the direction of the (mean) fluctuating stress $\tau$ acting on that segment, and we evaluate the mean square work:
\begin{equation}
	\langle W^2_{\tau}\rangle_{L,w} = \left\langle \left(\frac{1}{L} \int_0^L \langle \Delta E(x,w) \rangle_{\tau(x,0)}^2 \dx\right)^2 \right \rangle
	= \langle \tau^2 \rangle_L b^2 \left(\int_0^w \Phi_{\perp}(y/a) \dy\right)^2
\end{equation}
This expression contains the average of the square fluctuating stress over the length $L$. The result is given by (see Appendix):
\begin{equation}
	\langle \tau^2 \rangle_L = \langle \tau^2 \rangle \frac{1}{L} 
	\int_{-L/2}^{L/2} \Phi_{\parallel}\left(\frac{x-x'}{a}\right)\dx
	= \langle \tau^2 \rangle \frac{\xi - r(L)}{L}
\end{equation}
where $\xi = \int_{-\infty}^{\infty} \Phi_{\parallel}(x/a)\dx = \beta_1 a$ and $r(L) = 2\int_{L/2}^{\infty} \Phi_{\parallel}(x/a)\dx$. 
For sufficiently localized correlation functions one can, for $L \gg \xi$, neglect the residual $r(L)$. One then recovers the result for $\delta$-correlated fluctuations where $\Phi_{\parallel}(x) = \xi \delta(x)$.
We make this approximation in the following unless otherwise stated.  

Under the fluctuating stress, segments move to reduce their energy. The characteristic energy reduction for a segment of length $L$ moving over the distance $w$ can be estimated as $E_{\rm RF}(L,w) \approx \sqrt{\langle W_{\tau}^2\rangle_{L,w}}$. 
To maintain connectivity between adjacent segments (which in general displace in different directions), the dislocation has to elongate. We estimate the corresponding energy cost by the energy per unit length of a triangular bulge of total width $L$ and amplitude $w$: 
\begin{equation}
     E_{\rm LT}(L,w) \approx \frac{2 {\cal T} w^2}{L^2}
\end{equation}
The total energy change per unit line length for segments of length $L$ that displace independently to minimize their energy is then estimated as 
\begin{equation}
\Delta E(L,w) =  E_{\rm LT}(L,w) - E_{\rm RF}(L,w) \approx \frac{2T w^2}{L^2} - b \left(\left\langle \tau^2 \right\rangle \frac{\xi}{L}\right)^{1/2} \int_0^w \Phi_{\perp}(y/a) \dy
\end{equation}
The pinning energy per unit length derives by minimizing this expression
with respect to $L$ and $w$. Setting $\partial_w \Delta E(L,w) = \partial_L \Delta E(L,w) = 0$ gives 
\begin{eqnarray}
	0 &=& -4 \frac{{\cal T} w_{\rm p}^2}{L_{\rm p}^3} + \frac{b}{2} \sqrt{\left\langle \tau^2 \right\rangle \frac{\xi}{L_{\rm p}^3}} \int_0^{w_{\rm p}} \Phi_{\perp}(y/a) \dy
	\nonumber\\
	0 &=& 4 \frac{{\cal T} w_{\rm p}}{L_{\rm p}^2} - b \sqrt{\left\langle \tau^2 \right\rangle \frac{\xi}{L}} \Phi_{\perp}(w_{\rm p}/a) 
\end{eqnarray}
Both equations can be combined to eliminate the dislocation related parameters $({\cal T},b,L)$. It follows that the optimal displacement depends only on properties of the correlation function and obeys the equation:  
\begin{equation}
	2 w_{\rm p} \Phi_{\perp}(w_{\rm p}/a) = \int_0^{w_{\rm p}} \Phi_{\perp}(x/a)\dx
\end{equation}

We may now use these results to obtain from Eq. (7) the pinning length $L_{\rm p}$. With the notations $\xi = \beta_1 a, w_{\rm p}= \beta_2 a$ 
we get
\begin{equation}
	L_{\rm p} = \left(\frac{16\beta_2^2}{\beta_1
		\Phi_{\perp}^2(w_{\rm p}/a)}\right)^{1/3}\left(\frac{{\cal T}}{\sqrt{\langle\tau^2\rangle}b}\right)^{2/3} a^{1/3}
	\label{eq:lpin1}
\end{equation}
Setting finally ${\cal T} = \beta_0 \mu b^2$ we find the scaling relation
\begin{equation}
	L_{\rm p} = \left(\frac{16\beta_2^2\beta_0^2}{\beta_1 \Phi_{\perp}^2(w_{\rm p}/a)}\right)^{1/3}\left(\frac{\mu }{\sqrt{\langle\tau^2\rangle}}\right)^{2/3} b^{2/3}a^{1/3}
	= C_L \left(\frac{\mu }{\sqrt{\langle\tau^2\rangle}}\right)^{2/3} b^{2/3}a^{1/3}
	\label{eq:lpin2}
\end{equation}
Inserting (10) into (6) gives the pinning energy which results as:
\begin{equation}
	E_{\rm p} = -3 \left(\frac{\beta_1^2\beta_2^2 \Phi_{\perp}^4(w_{\rm p}/a)}{4\beta_0} \right)^{1/3}\frac{\sqrt{\langle \tau^2 \rangle}^{4/3}}{\mu^{1/3}} a^{4/3}b^{2/3}
	\label{eq:epin} = C_E \frac{\sqrt{\langle \tau^2 \rangle}^{4/3}}{\mu^{1/3}} a^{4/3}b^{2/3}
\end{equation}
Finally, the critical shear stress is estimated by equating the pinning energy to the work done by the external shear stress in moving the dislocation over the pinning displacement $w_{\rm p}$:
\begin{equation}
	\tau_{\rm ext,c} = \frac{E_{\rm p}}{b w_{\rm p}} 
	 = C_{\tau}\frac{\sqrt{\langle \tau^2 \rangle}^{4/3}}{\mu^{1/3}} \left(\frac{a}{b}\right)^{1/3} .
	\label{eq:fpinL}
\end{equation}

\section{Application to dislocation pinning in random alloys}

\subsection{Statistical properties of local shear stresses: A compilation of results} 

For an alloy constituting a random distribution of different atomic species which depending on atomic radius act as dilatation or compression centers in the effective medium of the alloy, statistical properties of the ensuing shear stress field were calculated by Geslin et. al. \cite{geslin2021microelasticityI,geslin2021microelasticityII}. In this paragraph we summarize their results. The mean square shear stress in an arbitrary plane is given by 
\begin{equation}
	\langle \tau^2 \rangle = \frac{V_{\rm a} \langle \Delta \epsilon^2 \rangle \mu}{30\pi^{3/2}a^3}
\end{equation}
where $V_{\rm a}$ is the atomic volume, $\langle\Delta\epsilon^2 \rangle$ is the mean square volumetric strain (dilatation or compression) introduced by an
individual atom into the effective medium constituted by the random alloy. The characteristic length $a$ arises in the treatment of Geslin et. al. as a regularization length that characterizes the distribution of the volumetric strain around the atom position. By comparison of theoretical results with stresses determined from molecular statics simulations, this parameter was
determined by Geslin et. al. for different alloys and found to be close to $a=1$\AA. 

The correlation functions in the shear direction ('longitudinal' correlation function $\Psi_{\rm L}$) and in perpendicular direction in the shear plane ('longitudinal' correlation function $\Psi_{\rm T}$) are
derived by Geslin et. al. as
\begin{eqnarray}
	\Psi_{\rm L}(u)&=&-\frac{30}{u^3} \left[\sqrt{\pi}\left(
	1-\frac{12}{u^2}\right)\erf\left(\frac{u}{2}\right)+\frac{1}{u}
	\left(12 + u^2\right) \exp\left(-\frac{u^2}{4} \right)\right],
	\nonumber\\
	\Psi_{\rm T}(u) &=& \frac{15}{u^3} \left[\sqrt{\pi}\left(
	1-\frac{6}{u^2}\right)\erf\left(\frac{u}{2}\right)+\frac{6}{u}
	\exp\left(-\frac{u^2}{4}\right)	\right],
\end{eqnarray}
where $u = d/a$ and $d$ denotes the distance from the origin in the respective direction. The correlation functions are plotted in Figure \ref{fig:edgecorrelation} which shows the anisotropic nature of the correlations. The corresponding correlation integrals, which are needed to evaluate pinning parameters, are given by 
\begin{eqnarray}
	\int_0^U\Psi_{\rm L}(u){\rm d}u &=&\frac{15}{U^4} \left[\sqrt{\pi}\left(
	U^2-6\right)\erf\left(\frac{U}{2}\right)+  6\exp\left(-\frac{U^2}{4} \right)\right],
	\nonumber\\
	\int_0^U\Psi_{\rm T}(u){\rm d}u &=& \frac{15}{8U^4} \left[\sqrt{\pi}\left(
	U^4-4U^2+12\right)\erf\left(\frac{U}{2}\right)+(2U^3-12U)
	\exp\left(-\frac{U^2}{4}\right)	\right],\nonumber\\
\end{eqnarray}
The correlation function in an arbitrary direction that makes an angle $\theta$ with the shearing direction is found to be 
\begin{equation}
\Psi(\theta,u) =  \Psi_{\rm L}(u) \cos^2\theta +  \Psi_{\rm T}(u) \sin^2\theta
\end{equation}
and the corresponding correlation integrals derive by analogous superposition
of the integrals in Eq. (15).

\subsection{Pinning of an edge dislocation}

We consider a perfect edge dislocation where the shear direction (the direction of the Burgers vector) is perpendicular to the dislocation line. We can thus identify the $x$ coordinate of the dislocation coordinate system with the transverse coordinate, $\Psi_{\rm T}(u) = \Phi_{\parallel}(x/a)$ and the $y$ coordinate with the longitudinal coordinate, $\Psi_{\rm L} = \Phi_{\perp}(y/a)$. 
\begin{figure}[tb]
	\centering
	\hbox{}
	\includegraphics[width=0.8\textwidth]{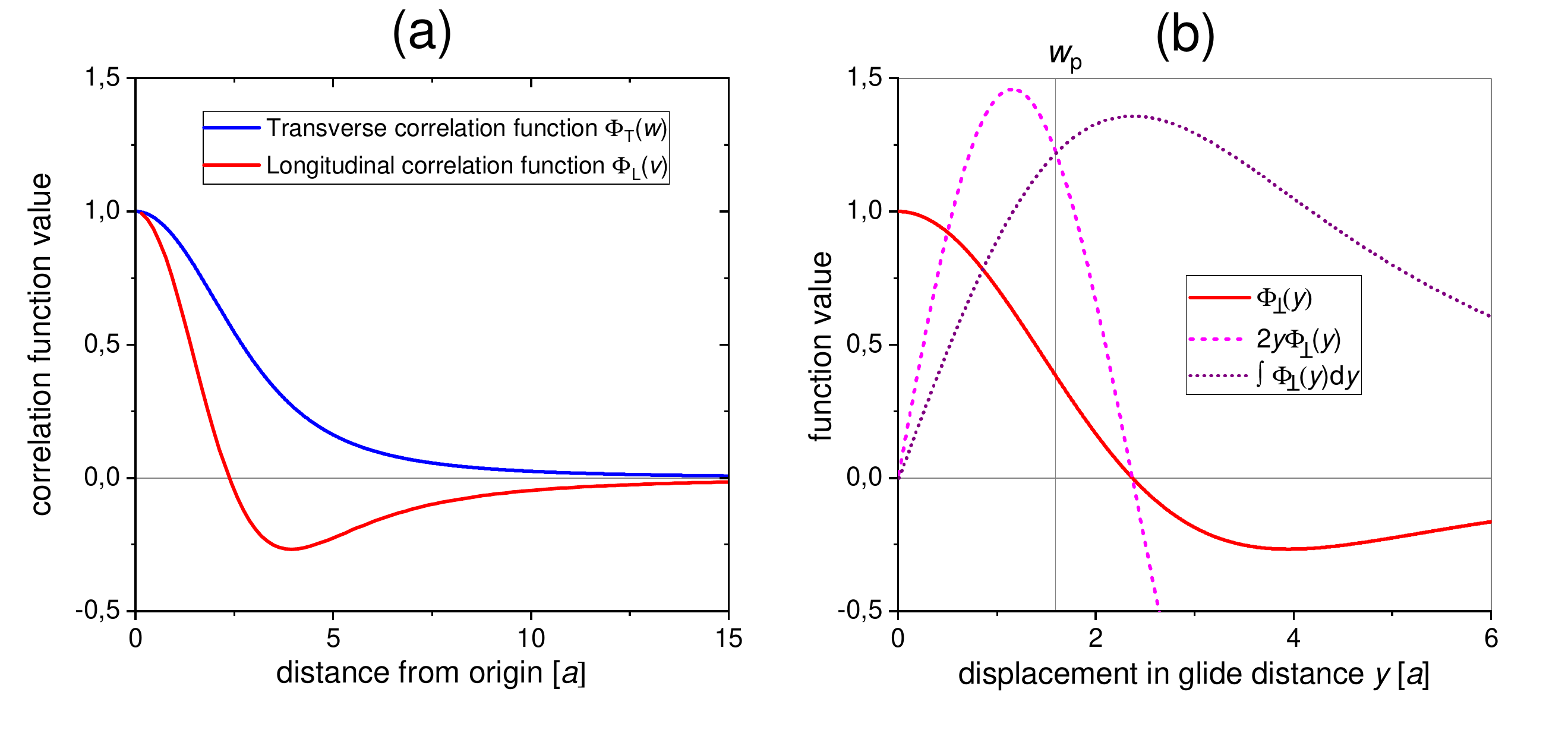}\hfill
	\caption{\label{fig:edgecorrelation}
		(a) Longitudinal and transverse correlation functions of the fluctuating shear stress in an effective medium with random volumetric misfit, blue: transverse correlations, red: logitudinal correlations, (b) determination of the optimal pinning displacement for an edge dislocation from the longitudinal correlation function.}
\end{figure}
The correlation length $\xi$ is evaluated as the integral of the parallel correlation function over the $x$ axis. From Eq. (15) it follows that 
\begin{equation}
\xi = \int_{-\infty}^{\infty} \Phi_{\parallel}(x/a) dx = \frac{15\sqrt{\pi}}{4} a \approx 6.6 a.
\end{equation}
Hence, the correlation length parameter $\beta_1 \approx 6.6$. 
In a simplified scaling analysis we neglect the residual that remains when the integral is restricted to the interval $|x| \le L/2$. For $L \gg \xi$ this residual can be evaluated from the asymptotic behavior of the correlation function as
\begin{equation}
	r = - 2 \int_{L/2}^{\infty} \Phi_{\parallel}(x/a) dx \approx 
	-30\sqrt{\pi} \int_{L/2}^{\infty} \left(\frac{a}{x}\right)^3 dx 
	= 60 \sqrt{\pi} a \left(\frac{a}{L}\right)^2.
\end{equation}
which provides, once the pinning length is determined, a simple check of the accuracy of the approximation $r \ll \xi$

We now use Eq. (8) to determine the optimum pinning displacement $w_{\rm p} = $ from the longitudinal correlation function $\Phi_{\perp}$ as shown in Figure \ref{fig:edgecorrelation}, (b). We find that $w_{\rm p} = 1.59 a$, hence $\beta_2 = 1.59$, and $\Phi_{\perp}(w_{\rm p}/a) = 0.38$. 

As a numerical example we consider a hypothetical Al$_{0.5}$Mg$_{0.5}$ fcc random solid solution for which simulations reported in Ref. \cite{geslin2021microelasticityII} give the values $\mu=20.7$ GPa, $a \approx 1$\AA,  and $\langle \tau^2 \rangle = 0.6$ GPa$^2$. For the Burgers vector length we use a value of $3.02$\AA,  which is the arithmetic mean of the Burgers vector length in Al and the $a$ lattice spacing in Mg. For the line tension associated with the bow out of an edge dislocation, we refer to Ref. \cite{zhai2019properties} who  determine, from thermal vibrations of an edge dislocation line, an effective line tension of ${\cal T} \approx 0.35 \mu b^2$, hence $\beta_0^{\rm e} = 0.35$. For a screw dislocation we consider a pre-factor $\beta_0^{\rm s} \approx 2.5 \beta_0^{\rm e}$. While isotropic elasticity theory predicts higher values, both experimental investigations \cite{mughrabi2001self} and recent atomistic simulations \cite{szajewski2015robust} suggest for fcc crystals a ratio $\beta_0^{\rm s}/\beta_0^{\rm e} \approx 2 \dots 2.5$, so using a ratio of 3 may serve as an acceptable compromise. More accurate and material specific values can, if needed, be deduced from molecular dynamics simulations of the thermal roughening of dislocation lines, using the method described in Ref. \cite{zhai2019properties}. All parameters we used in our numerical examples are compiled in Table 1. 

\begin{table}
	\centering
	\caption{Parameters used in numerical calculations.}
	\begin{tabular}{lll}
		\hline\noalign{\smallskip}
		Parameter& Unit& Numerical value\\
		\noalign{\smallskip}\hline\noalign{\smallskip}
		shear modulus $\mu$ & GPa & 20.7\\
		Burgers vector length $b$ & \AA & 3.05\\
		solute core parameter $a$ & \AA & 1.0\\
		mean square shear stress $\langle \tau^2 \rangle$ & GPa$^2$ &0.6\\
		line tension parameter $\beta_0^{\rm e}={\cal T}_{\rm e}/(\mu b^2)$ & -- & 0.35\\
		line tension parameter $\beta_0^{\rm s}={\cal T}_{\rm s}/(\mu b^2)$ & -- & 1.05\\
	\end{tabular}	
	\label{tabl:DSE}
\end{table} 

With these parameters we obtain for a pure edge dislocation $L_{\rm p} = 32.9$ \AA $\gg \xi = 6.18$ \AA. Eq. (11) gives for the pinning energy per unit length $E_{\rm p} \approx 1,83 \times 10^{-11} $J/m $\approx 0.01 \mu b^2$ , and the critical resolved shear stress derives from Eq. (12) as $\tau_{\rm ext,c}=390$MPa. 

With the pinning length given above we can estimate the residual $r(L_{\rm p})$ which with Eq. (17) follows as $r(L_{\rm p}) = 0.106$. This demonstrates that the approximation made in neglecting this residual is acceptable. We note that, in scaling analysis of pinning problems, this approximation is  often postulated {\em a priori} by approximating the correlation function in the line parallel direction as $\Phi_{\parallel}(x) = \xi \delta(x)$  The fact that we obtain, up to minor corrections, the same result as for such uncorrelated disorder demonstrates that an asymptotic third-order decay of the correlator, contrary to the conjecture of Ref. \cite{geslin2021microelasticityII}, not necessarily violates the assumption of short-range correlations used in standard scaling arguments. However, as we shall see, the case of a perfect screw dislocation illustrates how  standard scaling arguments can go wrong. 

\subsection{Pinning of a screw dislocation}

\begin{figure}[tb]
	\centering
	\hbox{}
	\includegraphics[width=0.9\textwidth]{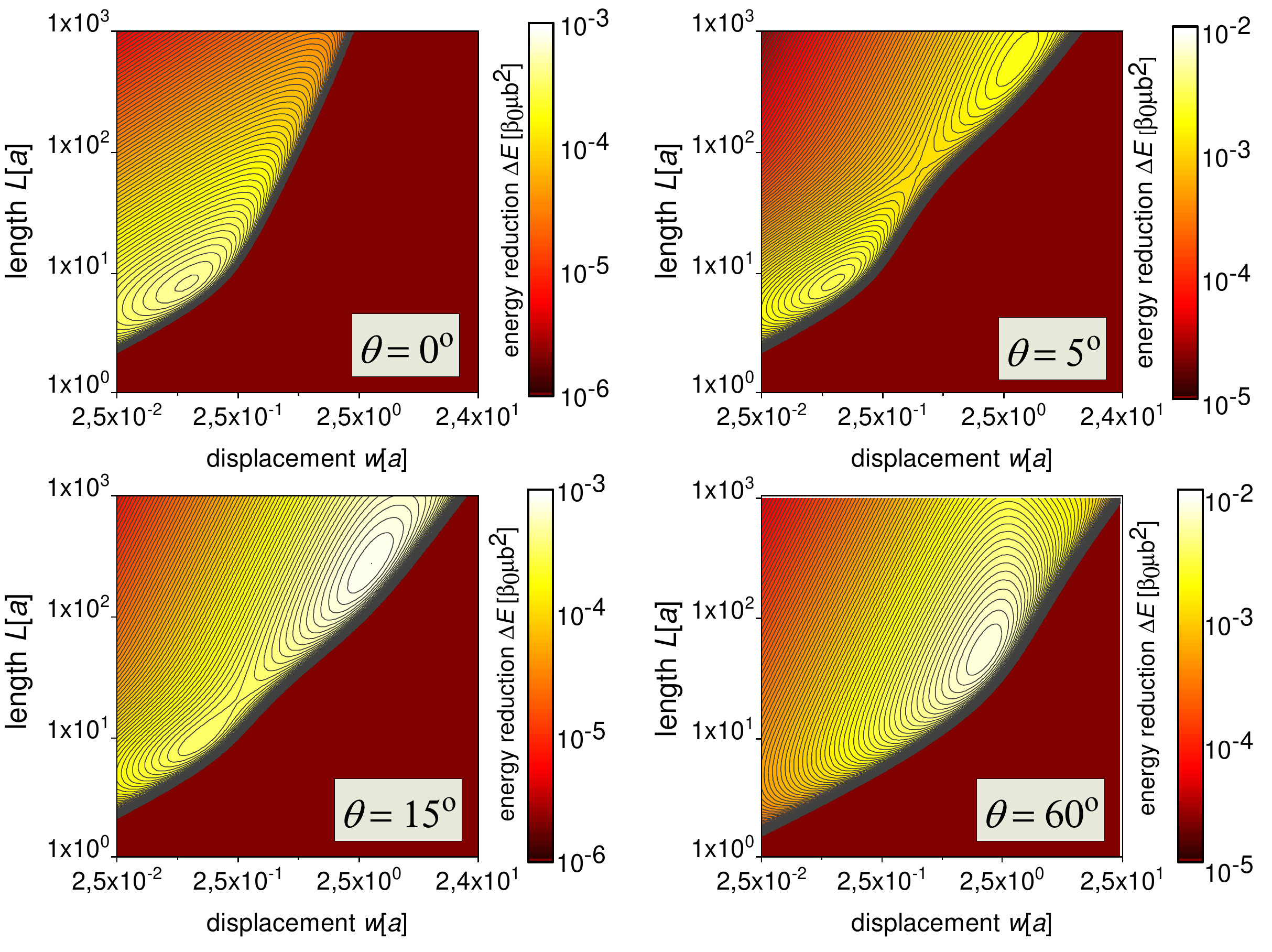}\hfill
	\caption{\label{fig:AngleEnergy}
		Energy landscapes for dislocations of various orientation, energy reduction of segments of length $L$ displacing over the distance $w$ 
		according to Eq. (19), angles are given in the figures.}
\end{figure}

If the dislocation orientation has screw orientation, the above treatment becomes spurious because the correlation function $\Phi_{\parallel}$ for a screw dislocation integrates to zero, the corresponding pinning length would therefore diverge and accordingly the pinning energy and pinning stress would be zero. For a screw dislocation we therefore investigate the energy gain $\Delta E(w,L)$ using the full expressions for the integral over $\Phi_{\parallel}$ that arises when averaging the fluctuating stress over a segment of length $L$:
\begin{equation}
	\Delta E(L,w) \approx \frac{2\beta_0 \mu b^2 w^2}{L^2} - b \left(\left\langle \tau^2 \right\rangle \frac{1}{L}\int_{-L/2}^{L/2} \Phi_{\parallel}(x/a) \dx\right)^{1/2} \int_0^w \Phi_{\perp}(y/a) \dy
\end{equation}
The corresponding energy landscape is illustrated in Figure \ref{fig:AngleEnergy}, top left. There is an energy minimum
located at $L \approx 6a$, in good agreement with the range over which the correlation function $\Psi_{L}$, i.e. the line parallel correlation function for screw dislocations, is positive. However, the corresponding optimal displacement $w \approx 0.08a$ is tiny and accordingly the maximum energy gain is small, amounting to $\Delta E_{\rm p,s} = 0.054\times 10^{-11}$ J/m $ = 2.85 \times 10^{-4} \mu b^2$. Nevertheless, because of the smallness of $w_{\rm p}$ Eq. (12) predicts a significant pinning stress of about 190 MPa, half the value for the edge dislocation. 

\subsection{Pinning of a general dislocation}

For a dislocation of general orientation $\theta$ where $\theta$ is the angle between the line direction and the Burgers vector, it follows from Eq. (15) that the line-parallel and line-perpendicular correlation functions are 
\begin{eqnarray}
	\Phi_{\parallel}(u) &=& \Psi_{\rm L}(u) \cos^2\theta +  \Psi_{\rm T}(u) \sin^2\theta,\nonumber\\
	\Phi_{\perp}(u) &=& \Psi_{\rm L}(u) \sin^2\theta +  \Psi_{\rm T}(u) \cos^2\theta.
\end{eqnarray}
Moreover, we assume that the line tension varies with the angle $\theta$ 
according to ${\cal T}(\theta) = \beta_0(\theta) \mu b^2$ where, for simplicity, we consider a simple sinusoidal variation:
\begin{equation}
	\beta_0 = \beta_0^{\rm s} \cos^2 \theta + \beta_0^{\rm e} \sin^2 \theta
\end{equation}
This allows us to study the angle dependence of the pinning parameters. 

\subsubsection{Edge-like pinning}

We first use a simplified treatment which, as in case of an edge dislocation, neglects the correlation residual $r(L)$. We thus base our analysis upon Eqs. (6-12). Since the longitudinal correlation function $\Phi_{\rm L}$ integrates to zero, the correlation length parameter is, upon neglecting the residual integral $r(L)$, given by \begin{equation}
	\beta_1 = \beta_1^{\rm e} \sin^2 \theta.
\end{equation}
Finally, we numerically establish the orientation dependence of the optimum pinning displacement $w_{\rm p}(\theta)$ and the corresponding function value
$\Psi_{\perp}(w_{\rm p}(\theta))$ using Eq. (8).

\begin{figure}[tb]
	\centering
	\hbox{}
	\includegraphics[width=0.8\textwidth]{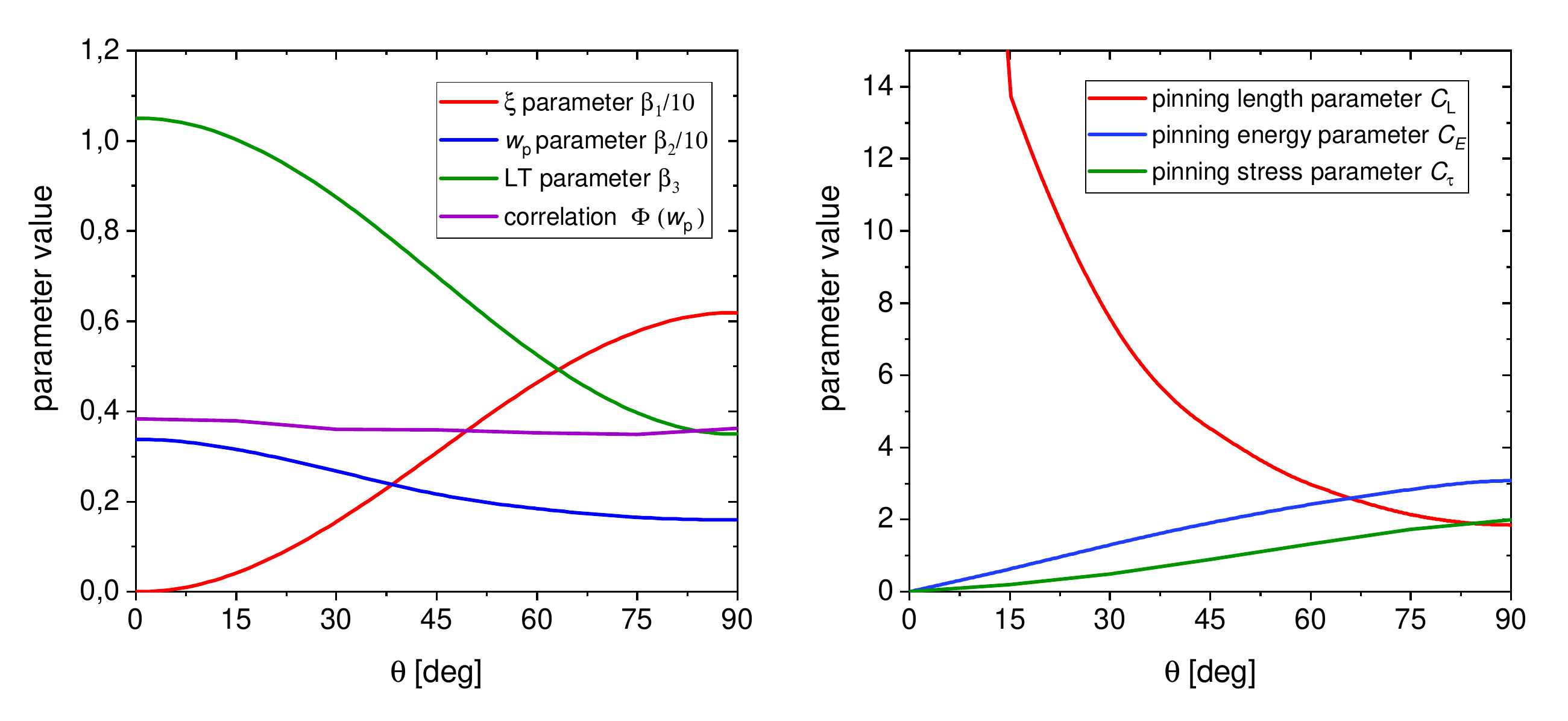}\hfill
	\caption{\label{fig:angleparameters}
		Left: Dependence of scaling parameters on dislocation orientation angle $\theta$; right: Corresponding parameters $C_{L}$, $C_E$ and $C_{\tau}$ which govern the angle dependence of the pinning length, pinning energy and pinning stress, respectively.}
\end{figure}

All $\theta$ dependent parameters are compiled in Figure \ref{fig:angleparameters}, left. 
From these parameters we can compute the angle dependent parameters $C_L, C_E$ and $C_{\theta}$. The respective physical variables are obtained from these by multiplication with material and fluctuation parametes according to 
Eqs. (10)-(12). The parameter $C_{L}$ increases as we move away from the edge orientation, whereas the correlation length parameter $\beta_1$ decreases. Accordingly, our analysis, which is built upon the smallness of $\xi/L$ and $a/L$, becomes more accurate. We denote this behavior as edge-like pinning.
The corresponding energy landscape is illustrated in Figure \ref{fig:AngleEnergy}, bottom right, for the case of a $60\deg$ dislocation.

In the screw dislocation limit our analysis however implies a diverging pinning length and vanishing pinning stress and energy, at variance with the findings of section 3.3. In fact, the energy minimum seen in Figure \ref{fig:AngleEnergy}, top left, for the screw dislocation actually represents a different class of 'screw-like' pinning behavior which we now investigate.  

\subsubsection{Screw-like pinning}

\begin{figure}[tb]
	\centering
	\hbox{}
	\includegraphics[width=0.8\textwidth]{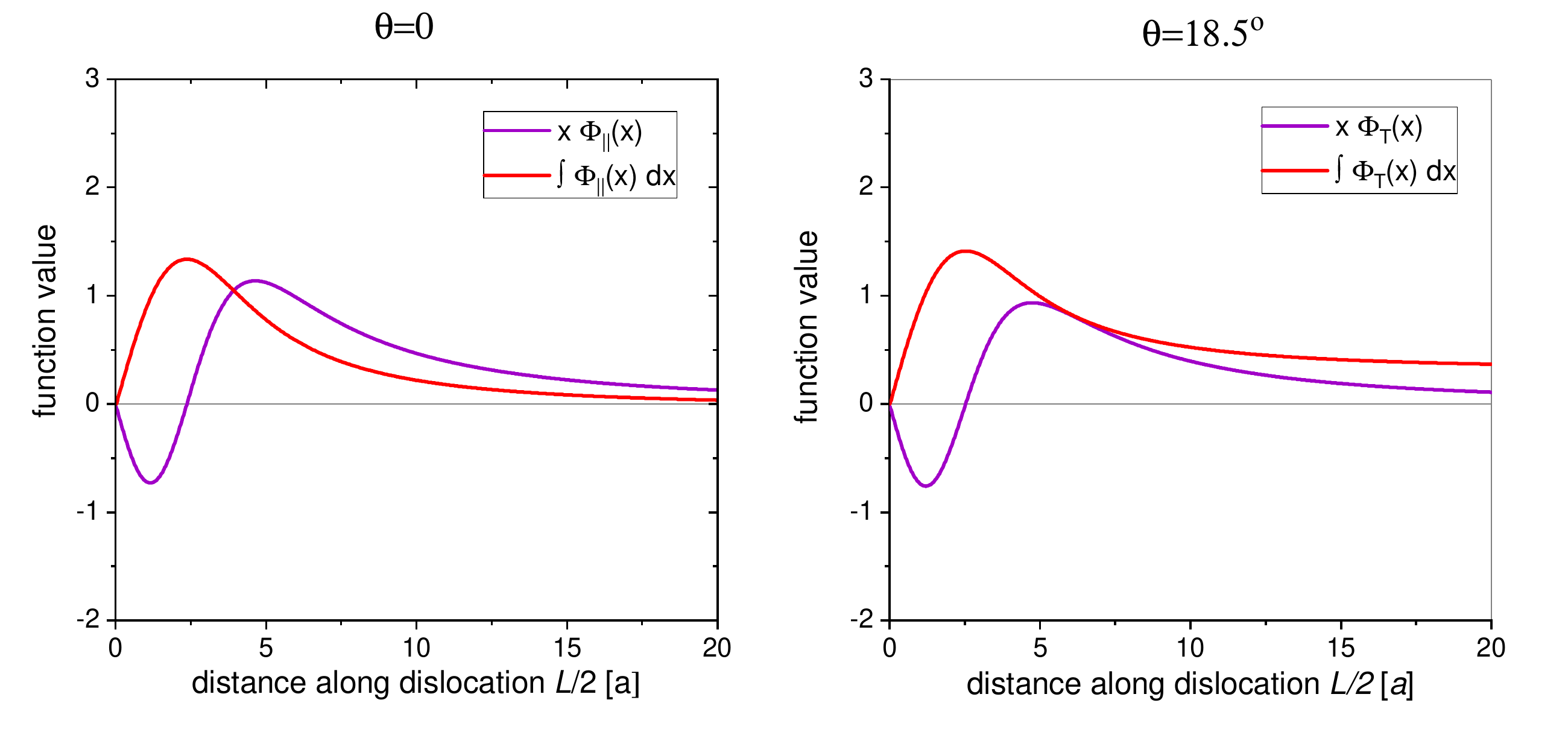}\hfill
	\caption{\label{fig:correlation}
		Determination of the pinning length for screw-like pinning for a pure screw (left) and a $30 \deg$ dislocation (right).}
\end{figure}

We base our analysis upon Eq. (19) which we simplify using the observation that, in case of screw dislocation pinning, the pinning displacement $w_{\rm p}$ which optimizes the energy gain is very small. Accordingly, we can approximate $\int_0^w \Phi_{\parallel}(x/a) \dx \approx w$. The extremum
condition $\partial_w \Delta E(L,w) = \partial_L \Delta E(L,w) = 0$ then gives 
\begin{eqnarray}
	0 &=& 4 \frac{{\cal T}w_{\rm p}}{L_{\rm p}^2} - b \sqrt{\left\langle \tau^2 \right\rangle}\left(  \frac{\int_{-L_{\rm p}/2}^{L_{\rm p}/2} \Phi_{\parallel}(x/a) \dx}{L_{\rm p}}\right)^{1/2}
	\nonumber\\
	0 &=& - 4 \frac{{\cal T} w_{\rm p}^2}{L_{\rm p}^3} - b \frac{\sqrt{\left\langle \tau^2 \right\rangle} w_{\rm p}}{2L_{\rm p}} \left( \frac{L_{\rm p}\Phi_{\parallel}(L_{\rm p}/2a)}{\left[\int_{-L_{\rm p}/2}^{L_{\rm p}/2} \Phi_{\parallel}(x/a) \dx\right]^{1/2}} 	
	- \left[\frac{\int_{-L_{\rm p}/2}^{L_{\rm p}/2} \Phi_{\parallel}(x/a) \dx}{L_{\rm p}}\right]^{1/2}\right)
	\nonumber\\
\end{eqnarray}
Both equations can again be combined to eliminate the dislocation related parameters $({\cal T},b,L)$, which produces an equation for the pinning length:
\begin{equation}
	- \frac{L_{\rm p}}{2a} \Phi_{\parallel}(L_{\rm p}/2a) = \int_{0}^{L_{\rm p}/2} \Phi_{\parallel}(x/a)\dx. 
\end{equation}
The situation is illustrated in Figure 4 for the cases of a pure screw ($\theta = 0$) and a 15 $\deg$ dislocation. In this range, the equation has two solutions, corresponding to a minimum and a saddle point of the energy surface. The two solutions merge at an angle slightly above $\theta = 18 \deg$, so there are no screw-like solutions for larger angles. At the same time, the saddle point moves to $L\to \infty$ as $\theta \to 0$ and thus merges with the edge-like minimum. The resulting energy landscapes are characterized by a ridge with two peaks (maxima of the energy reduction as the dislocation adjusts to the pinning landscape) and a saddle point. Such energy landscapes are illustrated in Figure \ref{fig:AngleEnergy}, top right and bottom left, for angles $\theta = 5 \deg$ and $\theta = 15 \deg$. 

\subsubsection{Numerical example}

We now apply the previously obtained relations to the case of the Mg$_{0,5}$Al$_{0.5}$ alloy studied by Geslin et. al. \cite{geslin2021microelasticityII}. We calculate pinning length, pinning energy and pinning stress for orientations between 0 and 90 $\deg$ using the relations for edge-like pinning (large pinning length, significant pinning displacement), and for orientations between 0 and 30 degrees, where the screw-like energy minimum exists, we compute the same parameters also for screw-like pinning. For the pinning length, we also show the unstable saddle point.

\begin{figure}[tb]
	\centering
	\hbox{}
	\includegraphics[width=0.99\textwidth]{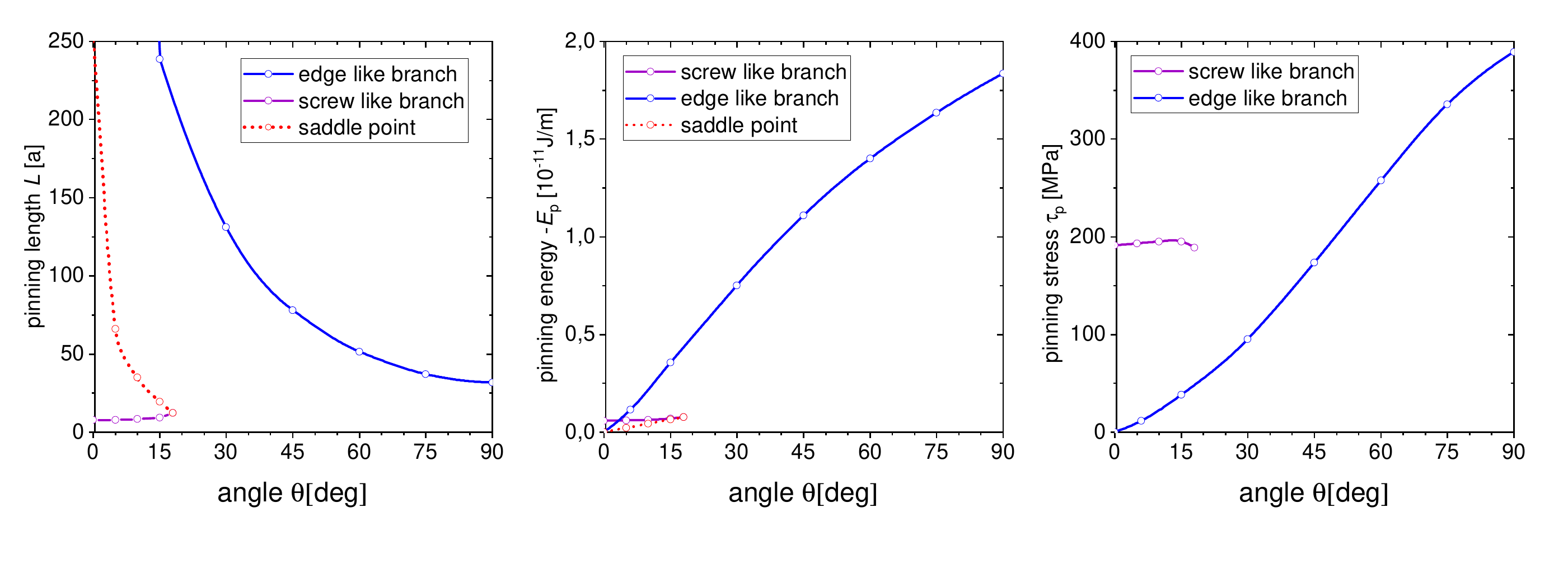}\hfill
	\caption{\label{fig:pinning}
		Pinning length (left), pinning energy (center) and pinning stress (right) as functions of dislocation orientation angle, calculated for an equiatomic MgAl solid solution with parameters given in Table 1. Blue lines denote parameters for edge-like energy minima, purple lines for screw-like minima, dotted red lines denote the values at the saddle point.}
\end{figure}

Parameters are found in Table 1, and results are compiled in Figure \ref{fig:pinning}. For near screw orientations the results show a co-existence of two pinning branches which show an interesting dichotomy, as the edge-like branch (except very close to the pure screw orientation) is characterized by an energy minimum that is deeper but much less steep, leading to a reduced pinning stress. The screw like branch, on the other hand, shows small pinning energies but, because the small pinning displacements lead to a steep slope of the energy landscape, appreciable pinning stresses. Which of these minima dominates the plastic behavior is difficult to decide without consideration of thermal activation and history effects.   

\section{Discussion and Conclusions}

Our calculations demonstrate that, in addition to well known line tension effects which lead to reduced pinning of screw relative to edge dislocations, the anisotropy of shear stresses created by an assembly of compression/dilatation centers has a major influence on the anisotropic pinning of dislocations. Edge dislocations and near edge dislocations follow the standard pinning paradigm despite a third law power decay of the stress correlator. Pinning of screw and near screw dislocations, on the other hand, is characteried by a different type of energy minimum with much reduced pinning length and pinning displacement. Pinning stresses of near-screw  dislocations are slightly lower than those of edge dislocations, but more importantly, the depth of the energy minimum for screw-like pinning is much lower than for edge-like pinning. As a consequence, at elevated temperature the motion of edge dislocations, being controlled by deep energy minima, is expected to be much more difficult than the motion of screw dislocations. This may serve as a generic explanation for recently reported observations in bcc High Entropy alloys \cite{kubilay2021high}. 

While we have established generic relationships that allow to separate the effects of line tension and of statistical parameters of the fluctuating stress field, in present form these relationships are applicable to perfect dislocations only. In High Entropy alloys, where stacking fault energies may be low, wide core splitting of dislocations introduces an additional degree of freedom which needs to be taken into account in pinning theories. Since the partials of a split dislocation have in general mixed character, we expect an even more complex energy landscape whose study we postpone to future work.

\section*{Declarations}
\subsection*{Competing interests}
  The authors declare that they have no competing interests.
\subsection*{Author's contributions}
MZ formulated the equations, R.W. performed numerical and analytical calculations, M.Z. drafted the manuscript which was edited and approved jointly by both authors. 
\subsection*{Funding}
M.Z. and R.W. acknowledge funding by DFG under Grant No. 1 Za 171/8-1. 
\subsection*{Availability of data and material}
Not applicable. 

\bibliographystyle{bmc-mathphys} 
\bibliography{references}   

\appendix
\section{Mean shear stress fluctuation acting on a straight dislocation segment}

The mean square fluctuations of the shear stress acting on a straight dislocation segment of length $L$ located between the points $(-L/2,0)$ and $(L/2,0)$ depend on the transverse correlation function. They derive as 
\begin{eqnarray}
	\langle \tau^2 \rangle_L &=& \langle \left(\frac{1}{L}\int_{-L/2}^{L/2} \tau(x,0)\dx\right)^2 \rangle = \frac{1}{L^2}\iint_{-L/2}^{L/2} \langle \tau(x,0)\tau(x',0)\rangle \dx\dx'\nonumber\\
	& = &\frac{\langle \tau^2 \rangle}{L^2}\iint_{-L/2}^{L/2} \Phi_{\parallel}\left(\frac{|x-x'|}{a}\right)\dx \dx'
\end{eqnarray} 
To evaluate the integral we introduce the notation $\tilde{x}=x-x'$:
\begin{equation}
	\langle \tau^2 \rangle_L = \langle \tau^2 \rangle \frac{1}{L^2}\int_{-L/2}^{L/2} \int_{-L/2-\tilde{x}}^{L/2-\tilde{x}} \Phi_{\parallel}(|\tilde{x}|/a) {\rm d}\tilde{x} \dx
\end{equation} 
The inner integral can be re-written as
\begin{eqnarray}
	\int_{-L/2-\tilde{x}}^{L/2-\tilde{x}} \Phi_{\parallel}(\tilde{x}) {\rm d}\tilde{x} &=& \frac{1}{2}\left( \int_{-L/2-\tilde{x}}^{L/2-\tilde{x}} \Phi_{\parallel}(\tilde{x}) {\rm d}\tilde{x}
	+ \int_{-L/2+\tilde{x}}^{L/2+\tilde{x}} \Phi_{\parallel}(\tilde{x}) {\rm d}\tilde{x} \right)\nonumber\\
	&=& \int_{L/2}^{L/2} \Phi_{\parallel}(\tilde{x}) {\rm d}\tilde{x} = \xi - r \quad,\quad r = 2 \int_{L/2}^{\infty}\Phi_{\parallel}(\tilde{x}) {\rm d} \tilde{x}.
\end{eqnarray} 
where we have exploited the symmetry of the correlation function.

\end{document}